# A Multilevel Scheduling Framework for Distributed Time-domain Large-area Sky Survey Telescope Array

Yajie Zhang[1,2], Ce Yu[1,2], Chao Sun[1,2], Zhaohui Shang[3], Yi Hu[3], Huiyu Zhi[1,2], Jinmao Yang[1,2], and Shanjiang Tang[1,2]
[1] College of Intelligence and Computing, Tianjin University, No.135 Yaguan Road, Haihe Education Park, Tianjin 300350, People's Republic of China
huyi.naoc@gmail.com
[2] Technical R&D Innovation Center, National Astronomical Data Center, No.135 Yaguan Road, Haihe Education Park, Tianjin 300350, People's Republic of China
[3] National Astronomical Observatories, Chinese Academy of Sciences, No.20 Datun Road, Chaoyang District, Beijing 100012, People's Republic of China

## Abstract

Telescope arrays are receiving increasing attention due to their promise of higher resource utilization, greater sky survey area, and higher frequency of full space-time monitoring than single telescopes. Compared with the ordinary coordinated operation of several telescopes, the new astronomical observation mode has an order of magnitude difference in the number of telescopes. It requires efficient coordinated observation by large-domain telescopes distributed at different sites. Coherent modeling of various abstract environmental constraints is essential for responding to multiple complex science goals. Also, due to competing science priorities and field visibility, how the telescope arrays are scheduled for observations can significantly affect observation efficiency. This paper proposes a multilevel scheduling model oriented toward the problem of telescope-array scheduling for time-domain surveys. A flexible framework is developed with basic functionality encapsulated in software components implemented on hierarchical architectures. An optimization metric is proposed to self-consistently weight contributions from time-varying observation conditions to maintain uniform coverage and efficient time utilization from a global perspective. The performance of the scheduler is evaluated through simulated instances. The experimental results show that our scheduling framework performs correctly and provides acceptable solutions considering the percentage of time allocation efficiency and sky coverage uniformity in a feasible amount of time. Using a generic version of the telescope-array scheduling framework, we also demonstrate its scalability and its potential to be applied to other astronomical applications.

*Unified Astronomy Thesaurus concepts:* Surveys (1671); Astronomical models (86); Astronomical methods (1043); Observational astronomy (1145)

## 1. Introduction

The development of astronomy has entered a new stage of comprehensive exploration of the dynamic universe. Wide-field, multiband, and high-cadence time-domain surveys are inevitable trends of next-generation of astronomy. At present, the time-domain survey projects represented by the Zwicky Transient Factory (ZTF; Bellm et al. 2018; Graham et al. 2019) and the Large Synoptic Survey Telescope (LSST; Ivezic et al. 2019) are based on single telescopes, for which it is difficult to take the coverage area, depth, and frequency of the survey into account. As a result, the time domain changes faster than the day scale and cannot be systematically explored. A telescope array, composed of telescopes of the same or different structures and functions in various geographical locations, has the potential to enable a novel observation modality and allow for complete large-scale time-domain sky surveys. The coordinated array observation of multiple telescopes in different sites can shorten the observation time interval, while maintaining the survey area and reducing the cost of the telescope (Liu et al. 2021). Therefore, a reliable, efficient, and unattended automatic scheduling system is a critical component to constructing a telescope array.

The astronomical observation scheduling is a complex problem (Dyer 2020). The scheduler has to deal with different kinds of constraints such as scientific goals, telescope performance, astronomical observation conditions, weather, opportunity cost, concurrent or consecutive use of resources, and observation quality feedback. Most of these constraints can change at any moment along with unexpected conditions that the telescopes need to work on together, making it more difficult to implement the observation scheduling. In addition, multitelescope-array scheduling is also a multiobjective problem as scientific goals can result in multiple optimization objectives in different aspects (Lampoudi et al. 2015).

Historically, the scheduling of individual, university, and institutional telescopes has been typically performed manually or semi-manually, which is infeasible in telescope arrays because of the decision complexity added by the increasing number of resources and observations. Furthermore, due to the need to study fast transient phenomena, near-real-time rescheduling responsiveness is required to achieve these scientific goals. Liu et al. (2018) developed a customized scheduler for the automatic AST3 survey, which is a priority queue-based scheduler considering both issues related to a general robotic telescope and the special conditions at Dome A. With the construction of large astronomical observation equipment and the consequent scientific requirements, more sophisticated schedulers have emerged. Especially noteworthy are the scheduling methods of the time-domain observation projects represented by LSST and ZTF. Naghib et al. (2019) modeled the scheduling problem of LSST as a Markovian decision process (MDP) and proposed a feature-based scheduler that does not rely on handcrafted observation proposals. It reduces the search space to a finite dimensional vector space by Markovian approximation. Bellm et al. (2019) presented a scheduling algorithm for wide-field imaging time-domain surveys and implemented it on the surveys of ZTF. They used an integer linear programming (ILP) solution, which can optimize an observing plan for an entire night by assigning targets to temporal blocks and the metric can weight contributions well from the time-varying airmass, seeing, and sky brightness.

The Las Cumbres Observatory Global Telescope (LCOGT)[4] is a robotic telescope network, and professional astronomers, citizen scientists, and educators can apply for access. For the telescope-array scheduling problem, the scheduler determines the facility used to observe a target (Lampoudi et al. 2015). Similar ILP techniques to that of ZTF are used to assign requested observations to telescopes. It is worth noting that the scheduling model of LCOGT differs from that of multiple geographically distributed telescopes forming an array to accomplish a specific sky survey project, as studied in this paper. In addition, it is a proposal-based scheduling, and the scientific merit of proposals is assessed by the Time Allocation Committee (TAC). Sol (2016) presented a scheduler solution based on mixed ILP (MILP) for the Atacama Large Millimeter Array (ALMA; Wootten 2003); time discretization and both static and dynamic constraints are considered in this scheduling model. However, the scheduling is also based on proposals and requires the TAC to assign the scientific priorities, which makes it challenging to schedule observations related to time-domain follow-ups (Alexander et al. 2017). The GLObal Robotic telescope Intelligent Array for e-Science (GLORIA; Castro-Tirado et al. 2014) aims for providing a global telescope array that citizens could use to conduct research in astronomy. A distributed scheduler that can coexist and interact with the local scheduler is deployed on the network. A probabilistic telescope decision algorithm proposed by López-Casado et al. (2019) decides which available telescope will be offered for the observation requests made by the users, and can adapt its response to changes in the array.

Large time-domain observations with distributed telescope arrays, foreseen to be available for ongoing and future surveys, brings a more challenging sequential decision problem and requires efficient scheduling methods and software solutions. The telescope array consists of multiple sites with different geographical locations, each of which may have various numbers of telescopes with different performance and observation statuses. Determining whether a field is observable is influenced by a combination of various factors that change over time.

To simplify, the scheduling of astronomical observations can be treated as a classical task allocation problem known as the job-shop problem (JSP), where $N$ ideal tasks are assigned to $M$ identical resources (Colome et al. 2012). Although there have been several attempts to apply various mathematical algorithms (linear programming, multiobjective evolutionary algorithms, reinforcement learning algorithms, etc.) for JSP problems (Chen & Tian 2019; Gao et al. 2019; Ghasemi et al. 2021), an elaborate method is needed to bridge the gap between the telescope-array community and the combinatorial optimization community. Also, compared with the classical JSP problem, our scheduling problem is more complex involving various time-varying constraints and multiple modes of task completion. In such context, this paper proposes a scheduling simulation and decision-making method for telescope-array observation. We present a multilevel scheduling model and implement a flexible framework of which the basic functionality is encapsulated into software components. In addition, our method studies the main constraints and objectives necessary to the distributed telescope-array scenarios, and identifies the main variables to analyze the performance of the proposed solution.

The reminder of this paper is organized as follows. In Section 2 we present the multilevel scheduling model that is proposed for the observation-scheduling problem of the telescope array. We propose the architecture of the scheduling framework and introduce the software implementation details in Section 3. Section 4 describes the scheduling algorithm and optimizations while Section 5 details the experiments and discusses the performance of the scheduler. Finally, conclusions are drawn and perspectives for future work are presented in the final section.

## 2. Model of Observation Scheduling for Telescope Array

### 2.1. Problem Definition

Because of the limitations of using a single telescope for the large-scale sky surveys stated in Section 1, we aim to use the distributed telescope array to complete high-cadence time-domain surveys under the condition of limited resources (telescopes and available observation time). Elaborate modeling and optimization algorithm are indispensable to obtain a reasonable observation arrangement order.

As shown in Figure 1, here a multilevel scheduling model is proposed to describe the optical time-domain sky survey observation process of the telescope array. The telescope-array observation-scheduling problem can be solved by the global scheduler and multiple site schedulers or telescope schedulers, respectively. First of all, the global scheduler is designed as a long-term scheduler, which acts as a central node to compute the scheduling strategies. It coordinates the telescopes of all observation sites and globally controls the progress of the sky survey and the utilization of resources. Periodically, it determines the suitable observation time of each site for each field according to the observation conditions and the image quality feedback. The preliminary scheduling results will be sent to astronomical observatories. At the same time, the global scheduler receives the execution feedback from the telescopes and makes decision changes to update and adjust the observation plans for the next scheduling coverage time for special cases (such as transient source tracking, telescope failure, poor observation quality, Internet connection down). Furthermore, the overall progress of the survey, that is, the completion of observation tasks, will be recorded and statistically analyzed, and important indicators will be visualized.

Notably, a large-scale time-domain sky survey is a dynamic process, so different observation modes may appear, the sky survey mode, follow-up observation mode, and so on. Telescopes with different filters may need to observe the same field at the same time for the follow-up of candidates with multiband photometry. In this case, telescopes located at the same site may need to be grouped to accomplish different tasks. Therefore, in addition to the site scheduler, a third-level telescope scheduler is necessary. The proposed site scheduler and telescope scheduler, as the scheduling decision maker for a short period of time, are responsible for allocating the tasks issued by the global scheduler to each telescope in the observation site, calculating the specific start and end times of observations with finer granularity and determining the final observation task queue for each telescope. During this process, the effects of real-time weather conditions around the observation sites, as well as the performance and status of telescopes are taken into account. In order to maximize the observation utilization of telescopes, the site/telescope schedulers can generate a few more redundant observation tasks in the task queue for each telescope. So when a field cannot be observed due to unexpected reasons, or temporary faults of the telescope exist, the site scheduler will first make a real-time response and decide the subsequent observation tasks of the current telescope, so as to reduce the impact of emergencies on the whole survey plan. Subsequently, the situation will be fed back to the global scheduler, which will make corresponding adjustments globally for follow-up task allocation, thus ensuring the mutual assistance and cooperation between telescopes in the array. Moreover, the observation quality of the field will be regularly fed back to the global scheduler. If poor observation quality occurs in some fields, the global scheduler will also increase the observation frequency of these fields when assigning observation considering the uniformity of the sky survey.

---
[4] https://lco.global/ https://lco.global/

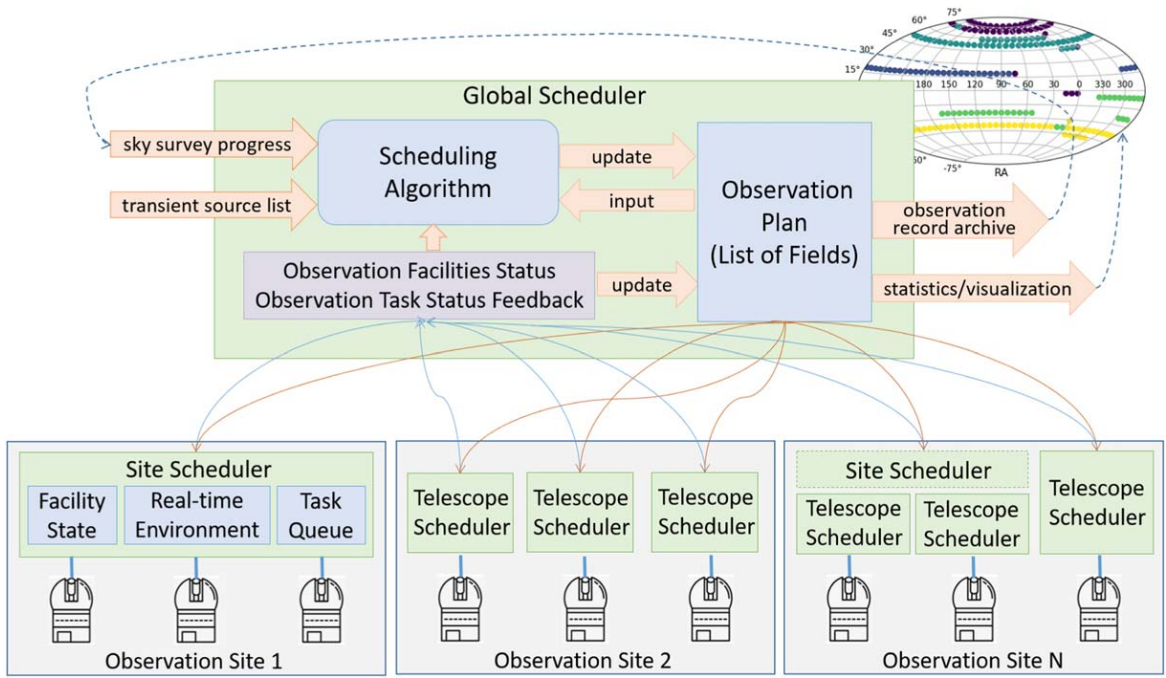

**Figure 1.** Multilevel model of the scheduling problem of the telescope array in the process of a time-domain large-scale sky survey. The overall survey plan is coordinated and controlled by the global scheduler and the site schedulers (and telescope schedulers if necessary) from different levels.

In general, making a decision for a visit at time $t$ for the telescope-array scheduling problem is mainly determined by the following factors:

1. The long-term mission-driven requirements, such as the uniformity of sky coverage and cadence, the minimization of the inactivity periods of telescopes, and the maximization of the quality of the images obtained in the observations.

2. The short-term observation-driven requirements, such as abrupt changes in the telescope performance and weather conditions at the observation sites.
3. The quality feedback of the images obtained by the telescopes following their observation tasks, for the overall science discovery of the survey.
4. The comprehensive execution preferences for the observation of crucial fields.

### 2.2. Scheduler Requirements

Therefore, in order to solve the above challenges and efficiently complete the telescope-array cooperative survey in a time-domain survey, a simulation and scheduling method oriented to the distributed telescope array is required. In order to realize the telescope-array cooperative survey model designed in Section 2.1 and accomplish multiple scientific objectives, the scheduling framework needs to meet the following requirements:

1. Simulation of generic survey observation modes, observation facilities and fields: This includes the simulation of telescope performance, state changes, the simulation and calculation of time-varying observation conditions, and the quality feedback from conducted observation tasks.
2. A feasible scheduling algorithm: The algorithm can synthesize all kinds of factors that affect the scheduling decision, solve the conflicts between different telescopes, and produce a solvable scheduling decision. Real-time scheduling is also a requirement; thus the speed of solving the optimization problem should be taken into account.
3. Dynamic adjustment of the multiple objectives of the survey in different modes: For such a complex multi-objective problem, different observation modes lead to different scientific objectives. The survey observation mode is to scan the ever-changing sky again and again as quickly as possible, which can make a movie of our universe and find huge amounts of variation sources (in particular dramatic events, such as mergers of dead stars, tidal disruption events, and supernovae). Meanwhile, the telescopes in the array can also enter the follow-up mode in which they are triggered by self and/or external alerts. In this mode, one or more telescopes will be scheduled to observe a target (may not be in the target list of the survey mode) with different observation configurations and constraints. So the scheduler may relax the observation constraints to increase the exposure time of the transient target. Manual or verbal intervention by experts should be minimized as much as possible.

4. Flexible response to unpredictable factors: The presence of unpredictable observation disruptions in the operation of the telescope array is mainly due to natural random processes (such as weather mutations) and unplanned instrument outages, which can be considered as random variables. Taking the above factors into account, the site scheduler needs to be able to respond quickly to unpredictable events, perform necessary rescheduling, and coordinate with other working telescopes to minimize the impact on the entire observation. Additionally, the global scheduler also needs additional logic to increase the perception of the array status and make planning adjustments when assigning follow-up observations.

5. Scalability: The scheduling framework needs to scale well with the expansion of the telescope array, which may include an increase in the number of observation sites and telescopes, or an increase in the considered time-varying constraints. Astronomers expect fine-grained simulations of the behaviors of observation sites and telescopes, which are subject to the curse of dimensionality (Bellman 1966). So the expansion of a telescope array may seriously degrade the performance of scheduling, which is a trade-off problem to be considered. Flexibility is also beneficial for further upgrade and substitution.

## 3. Implementation of Scheduling Framework

### 3.1. Architectural Design

The telescope-array observation-scheduling framework is based on the four-layer architecture demonstrated in Figure 2. The benefit of layering is to decouple the software by reducing the dependencies between layers, making it easier to extend functionality and adjust the implementation as requirements change.

The global layer is in charge of calculating the visibility and available observation time after the generation of the simulated locations of telescopes and fields. Considering the feasible time and feedback information from the site layer, the global layer is responsible for making a preliminary long-term scheduling analysis and decisions. The analysis consists of evaluating the observation condition constraints specified for each pair of observation site and field.

The site layer receives preliminary scheduling results from the global layer and is directly associated to each telescope in the array. However, the scheduling plans sent by the global scheduler could not be finally executed due to different issues: weather changes, technical problems of the telescope facilities, etc. In this case, the main function of the middle site layer is to perform the check for real-time dynamic constraints and generate the final observation tasks and the order of execution for telescopes. Meanwhile, the site scheduler is designed to reallocate the observation tasks for potential emergency situations. As the site layer establishes a direct communication with the telescopes, the telescope-array layer receives the specific tasks, and they are the ones that actually execute the observations.

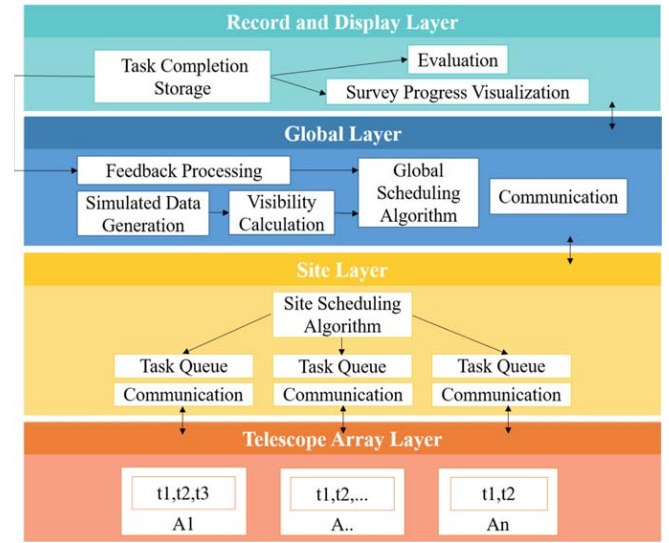

**Figure 2.** Architecture of the telescope-array observation scheduler based on the multilevel scheduling model. It is divided into the telescope-array layer, the site layer, the global layer, and the record and display layer. The global layer consists of simulation data preprocessing, global scheduling decision, and observation feedback processing. The site layer mainly includes the generation of a task queue for each telescope while the telescope-array layer is responsible for observation execution. The record and display layer contains the recording of task completion, visualization, and evaluation.

Once the observation has been executed, the acquired field image is available to astronomers. The execution of the task is recorded and evaluated by the record and display layer. Moreover, we provide the visualization function of the survey progress. By accessing the task completion and quality feedback of the images, the global scheduling algorithm may adjust subsequent scheduling decisions for global scientific goals.

### 3.2. Global Scheduler

The global scheduler is responsible for calculating the visibility of the observation field and for preliminary scheduling decision making during a long period of time. Notably, here we use the coordinates of the center of the observation field to represent its position and assume that, by pointing the telescope toward the center of the field, the visible field could be completely covered.

As stated in Section 1, one of the main complexities of the process of visibility calculation is that multiple types of factors are involved in the schedule decision of both the global scheduler and the site/telescope scheduler, which can also be viewed as input to the scheduling algorithm. Figure 3 shows the generic classification and examples of constraint inputs for the telescope-array observation-scheduling problem. Users can also supplement and customize constraint variables according to their specific telescope-array observation projects. According to the type of impact these factors have on the scheduling algorithm, we classify the input constraints for scheduling decisions into the following categories.

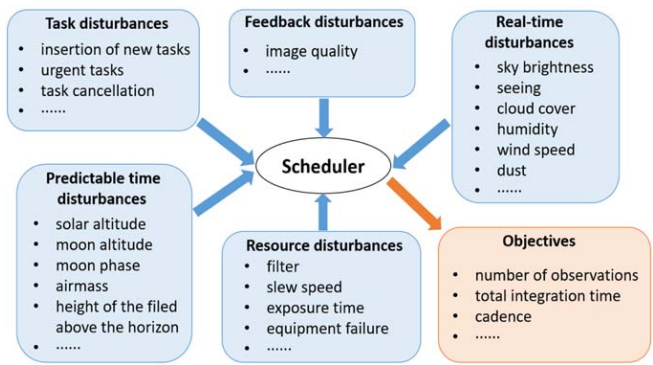

**Figure 3.** Simple representation of the input constraints and optimization objectives involved in the implementation of a telescope-array observation-scheduling algorithm. We propose to divide the constraints into five categories: task disturbances, predictable time disturbances, resource disturbances, real-time disturbances, and feedback disturbances. The scheduling algorithm also needs to support multiple science objectives.

1. Task disturbances: This is an external disturbance in the observation process, including new task insertion, task cancellation, and emergency tasks. The adjustment of the observation plans is mainly due to the short-term observations and tracking of special transient targets, which will affect the decision making of the global scheduler.
2. Predictable time disturbances: These are astronomical observation conditions (such as solar altitude, moon phase) and part of the weather effect that can be calculated and predicted, which have a direct effect on the visible time of the observed field. Meanwhile, it is also important to note that the airmass is one of the major obstacles for high-quality observation for a ground-based instrument. Filtering out the time of poor airmass conditions can provide higher-quality images and consequently result in more efficient operation of the observation. The predictable time disturbances are mainly considered by the global scheduling algorithm.
3. Resource disturbances: This mainly consists of the telescope performance, sudden equipment failures, and observation requirements such as camera filters, the amount of the time it takes to redirect the telescope and the dome to move from one target to the next, and exposure time. Fine-grained site scheduling and telescope scheduling are decided by the real-time resource features and changes in the proposed framework.
4. Real-time disturbances: All weather input data of the site/telescope scheduling algorithm are considered to be dynamic real-time variables. Long-term forecasts and statistics can be used, but unexpected changes in the variables require the schedulers to retrieve the current real-time weather conditions, which can widely vary from the forecast.
5. Feedback disturbances: When the quality of the observation results is not qualified or observation tasks are not completed due to unexpected circumstances, the global scheduler needs to reorganize the observation or dynamically adjust the follow-up plans from a global perspective.

Then the available time slots of each observation site for each observation field obtained after the visibility calculation process are input into the global scheduling algorithm, and the long-term observation plan is calculated by the scheduling algorithm. We call the output of the global scheduling algorithm "scheduling blocks," which can also be considered as the input of the site or telescope schedulers. The global scheduler calculates the available scheduling blocks for each site during the coverage time, and then hands them to the schedulers at the next level to assign specific tasks to the telescopes. The time slots of the scheduling blocks for different sites are different due to the difference in the station locations and observation conditions.

### 3.3. Site/Telescope Scheduler

During a specific short period of time, feasible scheduling blocks that are already arranged in the long-term phase are introduced to a new schedule phase, in which more real-time variables are considered. The dynamic variables and the equipment performance of the different telescopes of the site are taken into account. The main problem to be solved by the site or telescope scheduling algorithm is to determine the task execution queue for every telescope in a short time, which is a further refinement and local adjustment of the input scheduling blocks. In the absence of exceptional circumstances, we filter out telescopes that are in downtime and execute these tasks sequentially.

Notably, in our implementation, the constraints of real-time changes in the telescopes are not yet considered in detail. Accordingly our algorithm cannot be said to globally minimize slew overheads and the impact of real-time weather. Rather, the overall sky survey progress and uniformity can be guaranteed by roughly recording the uncompleted tasks of the sites and improving the attention and allocation of these fields in later scheduling. Such extensions are possible based on the current framework. Our framework will be extended to model the cost of slewing, simulate the short-timescale weather changes at the site, and conduct optimization for finer granularity. These extensions will introduce more complex uncertainty to the decision, and the trade-off between short-term efficiency and global optimization needs to be considered.

Moreover, our current scheduler implementation does not yet dynamically adapt to the changing of the observation mode in each site. We assume there is only a two-level scheduling, which means that the site scheduler at each observatory uniformly manages all telescopes. In practice, telescopes located at the same site may group and perform various real-time observing tasks like the follow-ups of the targets of opportunity (ToO) and the monitoring of periodic objects as well as optical transients. Thus how to concurrently handle those scientific cases is worth studying, although we think that a well-designed and flexible scheduling framework is a prerequisite to solve this problem, and is meaningful for further extension and reoptimization.

Finally, we have not attempted to optimize values related to the filters. As we are looking for a generic telescope-array observation-scheduling solution, rather than for a specific telescope-array observation project, it is assumed that the current scheduling result is the final result after adding all filter observations.

### 3.4. Dynamic Adjustment

In order to construct the cooperative observation of the telescope array, the proposed scheduling framework can dynamically adapt to observation condition changes, which mainly includes two parts: site adjustment and global adjustment. For cases where the telescope equipment is unable to perform the next observation task due to real-time dynamic variables, the site scheduler will quickly arrange other available telescopes to execute the remaining tasks for near-real-time responsiveness. This is because telescopes in the same site have the same available observation time slots in a short period of time regardless of dynamic factors. That is to say the scheduling blocks for each site are interchangeable. In this way, the missed observation tasks will then be reported to the global scheduler through the site scheduler for subsequent adjustment of the observation plan, which is beneficial for minimizing the global impact. In the meantime, the long-term global scheduler adapts its schedule map when the observation quality feedback is not satisfactory. We will explore the global rescheduling strategy in detail in a future work.

## 4. Scheduler Algorithm and Optimization

As we stated above, in the scheduling algorithm part this paper focuses more on the global scheduling level and the scalability of the method. As the core module of the global layer, the goal of the global scheduling algorithm is to generate a series of scheduling blocks for each observation site. As we described in Section 2.1, this problem can be treated as a multiobjective and multiresource optimization problem. Several optimizations can be made simultaneously in the observatory while multiple resources may be available concurrently, but they are not interchangeable due to their distinct locations. Thus, the global scheduling problem can be modeled as a distributed constraint satisfaction problem (distributed CSP; Yokoo et al. 1998), so several techniques can be used. This paper proposes a MILP (Floudas & Lin 2005) solution; the mathematical formulation is described below.

### 4.1. Parameters and Assumptions

Considering the previous assumptions and considerations, in the first place, a set of $n$ observation fields $F = \{F_1,...,F_n\}$ is defined to represent the regions of the sky that will be observed in the survey mode of the telescope array. The available observation sites are defined as a set of $r$ sites $\{S_1,...,S_r\}$, which can consist of various numbers of telescopes. An execution priority score $P_i$ is defined as a value that is associated to the need to increase the observation frequency of a certain field for some scientific purposes. Each field $F_i$ observed by each site $S_j$ is composed of a variable set of scheduling blocks $SB = \{SB_{ij1},...,SB_{ijs}\}$. The initial available set of scheduling blocks for each pair of field and site is obtained by calculating the observation conditions related to the long-term global scheduling mentioned in Section 3.2. Furthermore, we discretize time into slots, whose lengths can be uniform or nonuniform. A set of time slots represented by $T = \{T_1,...,T_l\}$ tells us when $SB_{ijs}$ can be executed during the time horizon covered by the global schedule.

### 4.2. Decision Variables

$Y_{ijst} = 1$ if the $SB_{ijs}$ of field $F_i$ and site $S_j$ is executed in the time slot $t$; 0 otherwise.

### 4.3. Objective Functions

To sufficiently respond to the variations in the design parameters, we design the objective function from the following aspects. First, in order to make full use of the time when the observation conditions are met, speed up the survey progress and improve the survey frequency of crucial observation regions, we prefer to maximize the sum of execution priorities of the scheduling blocks of fields that were successfully scheduled,

$$\max \sum_{i \in F} \sum_{j \in S} \sum_{s \in SB} P_i Y_{ijs}, \quad (1)$$

and maximize the sum of the scheduled observation lengths of the available time slots. In order to reduce the overhead of telescope switching and improve the observation utilization, a scheduled long-time block is preferred to a scheduled short time block.

$$\max \sum_{i \in F} \sum_{j \in S} \sum_{s \in SB} \text{length of}(SB_{ijs}) Y_{ijs}. \quad (2)$$

Furthermore, as illustrated in Figure 3, a large-scale time-domain survey requires maintaining the number of observations, total integration time, and cadence of fields as uniform as possible. The objectives of our method deliberately do not contain factors that account for these concerns, because they have no general quantitative relationship to our objective functions. Instead, we design a metric score of each scheduling block and try to maximize it as an objective to impose these constraints and optimize these metrics. The calculation formula of the metric score is as follows:

$$\text{score} = \rho * \frac{t\_c - t\_l}{f * T}. \quad (3)$$

The numerator means the time since a certain field was last observed ($t_c$ for the current time and $t_l$ for the end of the last observation time), with $\rho$ being an adjustable constant parameter. The smaller the value is, the closer the time when field was last accessed; thus it should be scheduled for observation as soon as possible. The denominator represents the total access frequencies of this field that has been observed times the total observation time, with $f$ and $T$ denoting the cumulative observation frequency and time, respectively. As our goal is to maximize the score value of each scheduling block, fields that were previously less observed are more likely to be selected for subsequent scheduling. Therefore, the uniformity requirements are well performed and optimized with respect to the changes in the parameters.

### 4.4. Constraints

The telescope array can only execute one SB at a specific time:

$$\sum_{i \in F} \sum_{j \in S} \sum_{s \in SB} X_{ijst} \leqslant 1 \quad \forall t \in \{1,...,l\}. \quad (4)$$

Each SB can be executed at most once:

$$\sum_{t \in T} Y_{ijst} \leqslant 1. \qquad (5)$$

In addition, depending on the observation project, multiple telescopes may be required to coordinate observations of the same field at the same time. In order to improve the efficiency of observations, while one telescope is observing a certain area, other telescopes may be assigned to perform different tasks. So our algorithm is flexible to add various constraints imposed by specific requirements.

## 5. Experiments and Evaluations

To evaluate the performance and capabilities of the scheduler, various experiments are performed. In this section, the experimental data sets and execution parameters are first introduced, followed by the scheduling decision results, which are presented by detailing the obtained results of the proposed performance metrics. Finally, the scheduling performance and scalability of the scheduling framework are discussed.

All implementation and experiments were developed on an Ubuntu server equipped with a four Intel Xeon CPU (four cores clocked at 2.2 GHz) and 32 GB memory. The scheduling algorithm was implemented in Python, which is publicly available[5] under an open-source license. We took advantage of a range of open-source Python libraries, including Astropy (Price-Whelan et al. 2018), Astroplan (Morris et al. 2018), Matplotlib (Hunter 2007), and Numpy (Van Der Walt et al. 2011). The framework Gurobi (Gurobi Optimization, Inc. 2014), a high-performance multipurpose solver, was used to implement the MILP optimization described in Section 4. It can provide native parallelization by initializing multiple candidate solutions on different threads and concurrently optimizing each, terminating when one thread obtains a solution. Moreover, we made attempts to make the scheduler interfaces telescope agnostic, which is beneficial for handling various astronomical observation projects and similar problems in other fields by simply replacing the model of the subproblem.

### 5.1. Data Sets and Execution Parameters

Considering that the distributed time-domain large-scale survey telescope array is still under development, there is a lack of a consistent volume of real data to analyze the proposed method. In such context, synthetical instances were generated looking forward to be as close as possible to real scheduling data. We used the positions of 20 global real observatories from the Astropy (Price-Whelan et al. 2018) library to simulate the distribution of observation sites in a distributed telescope array. Their details and geographical distribution are presented in Table 1 and Figure 4, respectively. In addition, 1000 positions of fields were generated for simulation whose distribution is shown in Figure 5. Part of them were derived from the coordinates of the discrete field grid in the ZTF project;[6] we also added some position coordinates for more comprehensive sky coverage.

The implementation of the proposed method has a wide variety of parameters used to change the behavior of the algorithm. Gurobi adds the possibility to modify internal parameters of the heuristics that are being used. Meanwhile, other parameters linked specifically to the algorithm can also be modified. Users can flexibly adjust the parameters by modifying the input configuration file. The main parameters used to execute all instances in default are detailed below:

1. Length of time slot: 5 minutes.
2. Coverage time of long-term scheduling: 24 hr.
3. Length of scheduling block: 30 minutes.
4. Input field file: data sets of size 50, 100, 200, 500, and 1000 are optional for evaluations.
5. Input site file: data sets of size 1, 2, 5, 10, and 20 are optional for evaluations.

---

[5] https://github.com/Melody888Evan/Telescope_Array_Observation_Scheduler.
[6] See https://github.com/ZwickyTransientFacility/ztf_sim/blob/master/data.

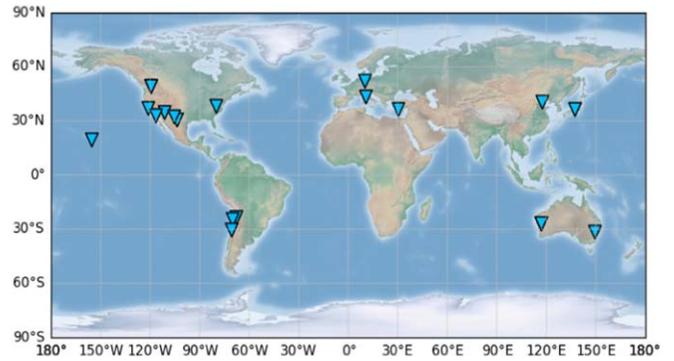

**Figure 4.** Geographical distribution of the observation sites used in the experiments. We select 20 different observatories or telescopes from the Astropy package (Price-Whelan et al. 2018) to simulate the locations of the sites contained in a telescope array, which are marked by triangles.

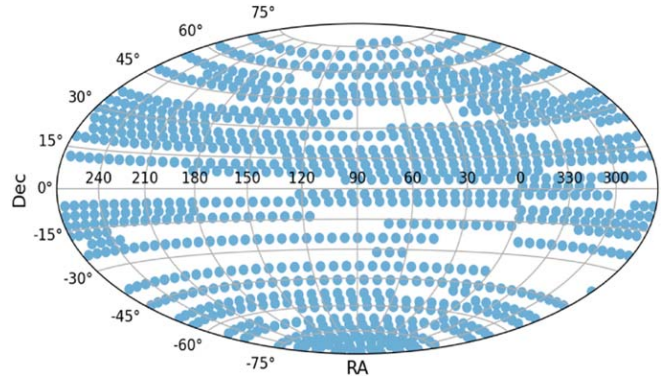

**Figure 5.** Spatial distribution of the simulated field data set of size 1000. These are the sky regions where scheduled observations are needed in the sky survey mode of the telescope array. The R.A. range is [0°, 358°.48], while the range in decl. is [−89°.05, 76°.55].

**Table 1**
Summary of Observation Site Information Used in the Experiments to Simulate the Composition of the Telescope Array

| Site | R.A. (deg) | decl. (deg) | Altitude (m) |
|---|---|---|---|
| Subaru | −155.48 | 19.83 | 4139.00 |
| TUG | 30.34 | 36.82 | 2500.00 |
| KAGRA | 137.31 | 36.41 | 414.18 |
| ALMA | −67.76 | −23.03 | 5000.00 |
| MWA | 116.67 | −26.70 | 377.83 |
| Cerro Tololo | −70.82 | −30.17 | 2215.00 |
| VIRGO | 10.50 | 43.63 | 51.88 |
| DRAO | −119.62 | 49.32 | 546.57 |
| G1 | 9.81 | 52.25 | 114.43 |
| CHIME | −119.62 | 49.32 | 555.37 |
| Sunspot | −105.82 | 32.79 | 2800.00 |
| Hale Telescope | −116.86 | 33.36 | 1706.00 |
| Lick Observatory | −121.64 | 37.34 | 1290.00 |
| Discovery Channel Telescope | −111.42 | 34.74 | 2337.00 |
| Murchison Wide-field Array | 116.67 | −26.70 | 377.83 |
| Green Bank Telescope | −79.84 | 38.43 | 807.00 |
| Beijing Xing Long Observatory | 117.58 | 40.39 | 950.00 |
| Anglo-Australian Observatory | 149.07 | −31.28 | 1164.00 |
| McDonald Observatory | −104.02 | 30.67 | 2075.00 |
| Paranal Observatory | −70.40 | −24.63 | 2669.00 |

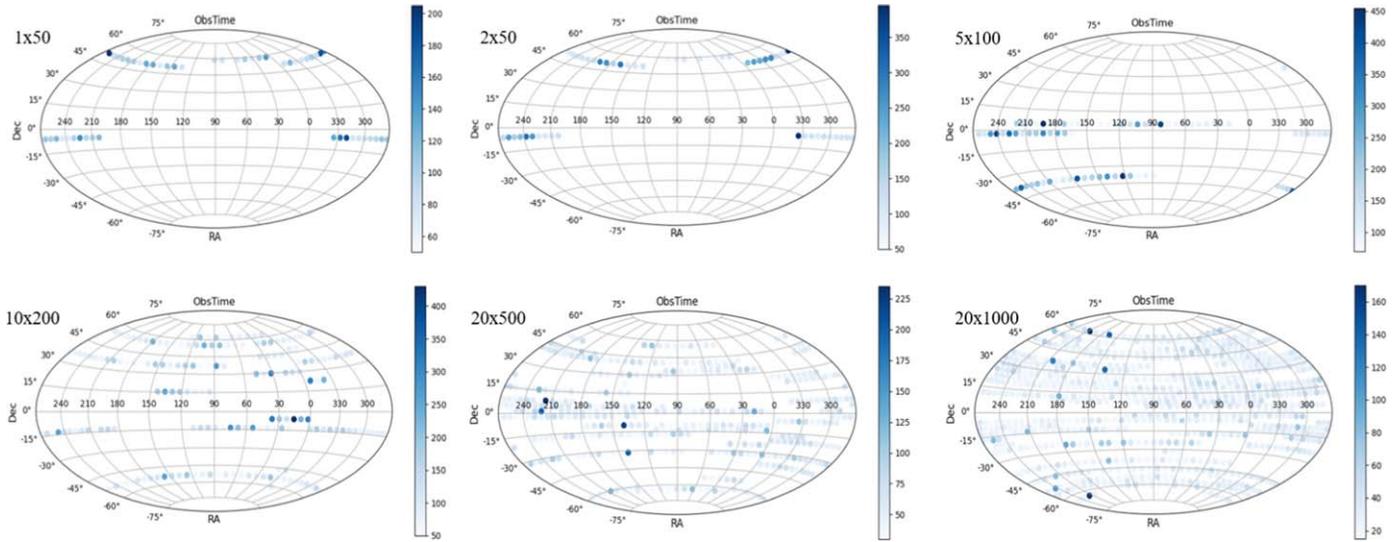

**Figure 6.** Coverage of cumulative observation time (in minutes) for various data sets after the duration of 10 long-term scheduling processes (10 days). According to the objectives of large-scale time-domain survey missions, the scheduler needs to provide a uniform coverage of the visible sky within each field. It can be seen that, with the expansion of the scale of the sky survey, the uniformity performance will be better.

### 5.2. Obtained Results and Discussions

For a multiobjective sky survey, comparing the overall performance of the different scheduling methods is quite challenging, especially because of the large number of competing factors that are involved in the scheduling performance evaluation. According to the scientific goals of the large-scale time-domain survey, the time utilization and the uniformity of observation in each field are two essential parts. So we propose to measure the observation quality of a scheduler from the following four aspects:

1. UNO: uniformity of the number of observations. We calculate the standard deviation of the number of observations for each field in the scheduling results to represent this value.
2. UCOT: uniformity of the cumulative observation time. Again, the standard deviation of the total observation time for each field is used.
3. UC: uniformity of cadence. According to the scheduling results, the time interval between the start of each scheduled observation and the end of the previous observation is calculated for each field, and the standard deviation is calculated to represent the uniformity between fields.
4. TAR: time allocated ratio. It is defined as the proportion of the time successfully allocated by the scheduler to all available observation times calculated under the observation conditions.

As shown in Figure 6, the proposed scheduler offers smoother time coverage as the scale of the survey increases. The more stable uniformity for larger-scale surveys is due to our fine discretization of the observable time slots and the practical design of the metric score in the decision-making stage. However, there are still a few regions that have more coverage time, which is mainly because of the potential competition between locations and observation conditions of the fields. Table 2 contains details of the scheduling results for combinations of different numbers of sites and fields, representing different sizes of telescope arrays. Generally, the uniformity of observations performs better with the progress of the sky survey while the overall time allocated ratio decreases. Comparing the values of different inputs especially when the survey only contains 50 or 100 fields, the UC and TAR show that there is a trade-off between these two values. While larger-scale scheduling offers better uniformity, the time utilization decreases around 40%–50%. In other words, the scheduler sacrifices the perfect time allocated ratio for a limited area of the sky to maintain a uniform observation for a larger scale of fields. Fortunately, when the size of fields is more than 200, this situation is significantly improved. The time allocated ratio reaches more than 90% and remains relatively stable. It reveals that the proposed scheduler is able to dynamically adjust the scheduling strategy according to the survey progress, so as to achieve stable sky coverage and efficient time utilization, which is consistent with our design.

Due to the differences in the astronomical phenomena in various seasons, the suitable time for observation varies from one field to another. The visibility of fields may face large differences with time. Therefore, we selected five sites and 100 fields and conducted scheduling simulation experiments over a time span of one year. For experimental efficiency, we adopt a sampling approach, sampling once a week, to represent the observations of this week. As shown in Figure 7, the uniformity is also relatively stable under the annual span. The intensity of observations is basically seasonally varied, which means that different time periods mainly focus on fields at different locations. Thus, it can be shown that with seasonal changes in astronomical observation

Table 2
Summary of Results of All Data Sets After the Duration of 1–10 Long-term Scheduling Processes

| Size of Sites | Size of Fields | | Size of Long-term Observations | | | | | | | | | |
|---|---|---|---|---|---|---|---|---|---|---|---|---|
| | | | 1 | 2 | 3 | 4 | 5 | 6 | 7 | 8 | 9 | 10 |
| 1 | 50 | UNO | 3.82 | 4.68 | 6.56 | 5.01 | 4.72 | 7.06 | 7.48 | 7.14 | 6.77 | 6.53 |
| | | UCOT | 19.12 | 23.39 | 32.82 | 25.04 | 23.58 | 35.28 | 37.42 | 35.71 | 33.85 | 32.65 |
| | | UC | 139.25 | 128.03 | 114.82 | 103.88 | 107.28 | 100.16 | 94.46 | 92.95 | 91.77 | 92.18 |
| | | TAR | 93.60% | 64.95% | 58.63% | 57.08% | 49.74% | 50.60% | 51.08% | 47.75% | 44.81% | 42.38% |
| 2 | | UNO | 5.23 | 4.79 | 4.45 | 4.19 | 9.63 | 11.90 | 14.40 | 17.94 | 17.58 | 16.94 |
| | | UCOT | 25.17 | 23.93 | 22.25 | 20.93 | 48.14 | 59.51 | 71.99 | 89.71 | 87.89 | 84.71 |
| | | UC | 101.20 | 98.16 | 95.72 | 94.65 | 85.37 | 79.90 | 75.65 | 70.92 | 70.76 | 70.31 |
| | | TAR | 94.56% | 67.26% | 53.49% | 49.09% | 50.50% | 49.00% | 47.33% | 47.97% | 45.68% | 44.39% |
| 5 | 100 | UNO | 4.46 | 11.15 | 12.62 | 14.25 | 16.77 | 17.72 | 19.40 | 18.79 | 18.25 | 19.72 |
| | | UCOT | 22.32 | 55.76 | 63.11 | 71.24 | 83.86 | 88.59 | 97.02 | 93.93 | 91.24 | 98.62 |
| | | UC | 156.31 | 131.84 | 115.14 | 102.88 | 95.94 | 89.77 | 86.48 | 86.06 | 85.85 | 82.37 |
| | | TAR | 92.74% | 79.07% | 73.66% | 71.32% | 67.64% | 66.04% | 61.50% | 59.37% | 57.57% | 56.55% |
| 10 | 200 | UNO | 2.46 | 3.73 | 4.94 | 6.04 | 7.72 | 9.22 | 9.92 | 12.39 | 12.61 | 13.15 |
| | | UCOT | 12.28 | 18.63 | 24.69 | 30.19 | 38.60 | 46.10 | 49.61 | 61.94 | 63.04 | 65.73 |
| | | UC | 199.64 | 170.99 | 150.30 | 137.31 | 126.59 | 119.04 | 112.97 | 107.99 | 103.51 | 99.57 |
| | | TAR | 96.12% | 91.37% | 89.54% | 87.57% | 85.58% | 83.93% | 82.36% | 79.88% | 78.93% | 77.41% |
| 20 | 500 | UNO | 1.28 | 1.87 | 2.39 | 3.22 | 3.11 | 3.95 | 3.84 | 5.15 | 4.94 | 4.85 |
| | | UCOT | 6.38 | 9.36 | 11.95 | 16.11 | 15.56 | 19.73 | 19.21 | 25.73 | 24.72 | 24.24 |
| | | UC | 194.43 | 168.77 | 148.27 | 131.42 | 122.81 | 109.15 | 101.88 | 91.68 | 82.85 | 76.48 |
| | | TAR | 95.98% | 95.21% | 94.51% | 91.98% | 90.69% | 89.41% | 88.51% | 87.14% | 86.59% | 86.65% |
| | 1000 | UNO | 0.85 | 1.08 | 1.35 | 1.52 | 1.71 | 1.95 | 2.59 | 2.59 | 3.09 | 2.98 |
| | | UCOT | 4.27 | 5.41 | 6.76 | 7.60 | 8.56 | 9.75 | 12.97 | 12.96 | 15.45 | 14.90 |
| | | UC | 187.32 | 178.41 | 172.35 | 164.30 | 150.46 | 136.19 | 120.91 | 108.49 | 99.65 | 87.68 |
| | | TAR | 98.22% | 98.19% | 97.83% | 96.92% | 95.93% | 94.85% | 94.44% | 94.31% | 92.96% | 93.18% |

**Note.** The scheduling results are evaluated from four metrics: UNO, UCOT, UC, and TAR, as defined in Section 5.2.

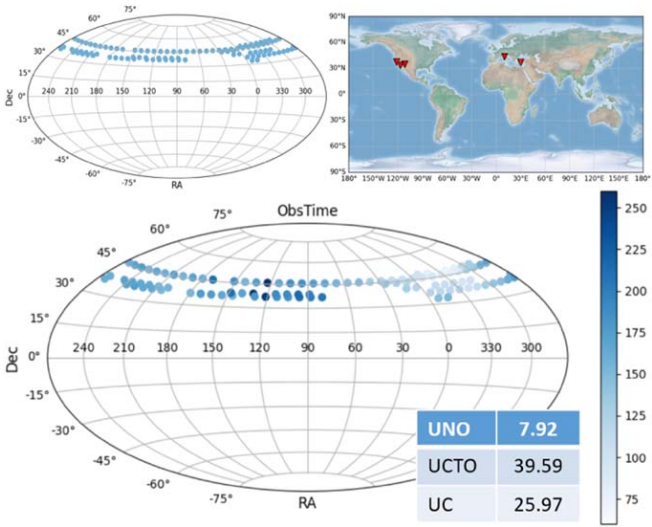

**Figure 7.** Observation uniformity evaluations over a time span of one year. The first row represents the locations of the input five sites and 100 fields. The second row shows the results of coverage of cumulative observation time (in minutes) and the uniformity metrics. The annual uniformity of observations shows a stable behavior.

conditions, our scheduler can dynamically adjust the task arrangement to maintain global uniformity.

### 5.3. Scheduling Performance

To evaluate the performance of the proposed scheduling method, the total execution time is divided into the preprocessing time and the global scheduling time. Notably, we focus on the scheduling calculation time of the method, so the specific scheduling control time of the site for each telescope is not included. In the preprocessing process, it is necessary to calculate each field and site within the next scheduling time separately to obtain the feasible time window according to the observation conditions. So this process is computationally complex and takes quasi-linearly more time as the amount of data increases. As shown in Figure 8, the experimental results are consistent with our analysis. In the case of 20 sites cooperating to observe 1000 fields, the preprocessing time of the observation conditions accounts for approximately 60% due to the complex multidimensional features generated by the large quantity of input sites and fields.

In addition, as for the global scheduler, it takes approximately 40%–55% of the total execution time and also grows quasi-linearly as the amount of data increases. Because with the simultaneous increase in the number of observable fields, the decision maker will have more choice space when making scheduling decisions, and the calculation time will correspondingly be longer. However, the speedup ratio increases steadily as the data volume increases, which shows good scalability. For long-term and large-scale scheduling (length of 24 hr in the experiments), the decision time for 1000 fields is less than 1 hr and 2 hr for 10 sites and 20 sites, respectively. This indicates that the performance of our method is acceptable and sufficient for the subsequent dispatch of telescopes in real-life usage.

### 5.4. Airmass Distributions and Limiting Magnitudes

As the site distribution and field distribution can affect the observation conditions (seeing, sky brightness, etc.), and therefore affect the observation quality, in order to make a comprehensive exploration of the various influencing factors, the above experiment does not make too put observation quality constraints on the observable time. In that case, to make the simulation more close to the requirements of the real-time-domain survey projects, three scenarios are designed, of which the observation height angle is not limited or is limited to more than 30° and more than 60°. Therefore, the visible time is obtained by prefiltering out unsuitable observation conditions and quality. By passing the airmass limit to the optimization metric, we encourage the visiting of fields with better observation quality.

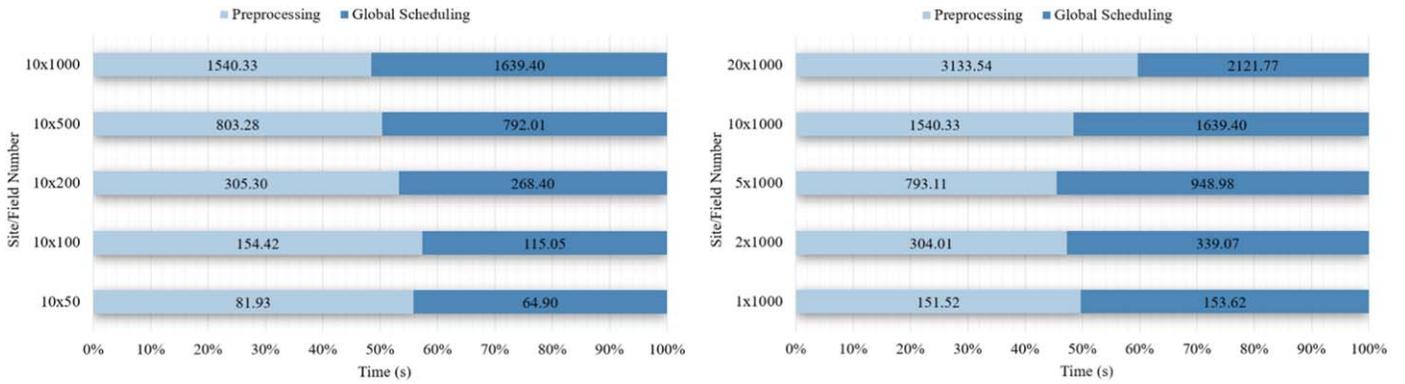

**Figure 8.** Scheduling performance evaluations on different number of sites and fields. Left: data sets of 10 sites and 50, 100, 200, 500, and 1000 fields, respectively. Right: data sets of 1000 fields and 1, 2, 5, 10, and 20 sites, respectively. With the increase in the data volume, the preprocessing time and global scheduling time increase almost linearly. Meanwhile, the time consumption ratio is also affected by the observation conditions.

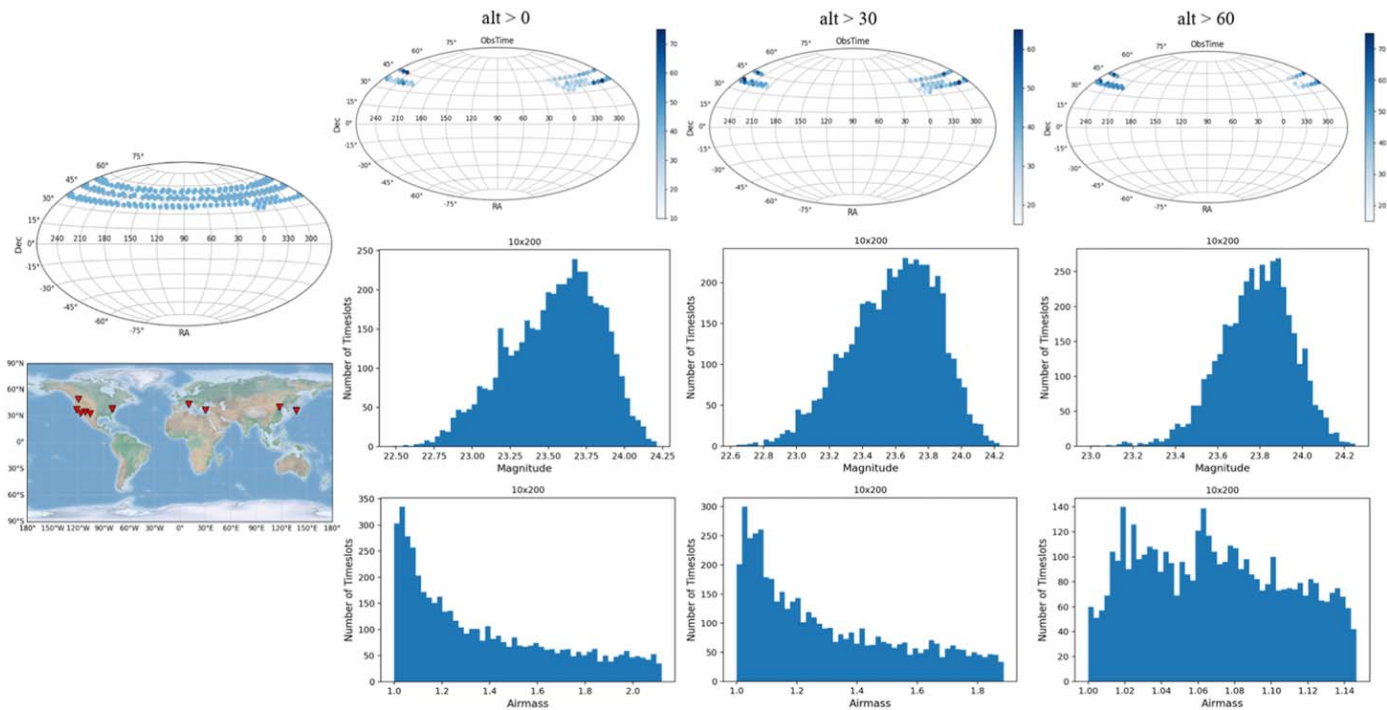

**Figure 9.** Coverage of cumulative observation time distributions and histograms of $5\sigma$ limiting magnitudes of the input 10 sites and 200 fields under various alt constraints after the duration of 10 long-term scheduling processes. The first column shows the locations of input sites and fields. Fields and times with poor observation quality are removed. Moreover, the fields around the zenith will be prioritized.

The median and standard deviation of the airmass are calculated to reflect the overall quality of the collected data by the telescope array. Simultaneously, an open-source library[7] was utilized to evaluate the $5\sigma$ limiting magnitude for the scheduled observations. In all experiments we use the filter "z" to calculate the magnitude. Note that it calculates the limiting magnitude of LSST, and the results will be different for different telescopes. Similar models can be used for the calculation of the observation conditions in the telescope-array scheduling problem, but the performance problem of large-scale calculations can be one of the most serious bottlenecks. The results obtained are shown in Figure 9 and Table 3. Under various scenarios, the results of the four proposed metrics and observation quality results are relatively stable. This reveals that the scheduler can provide stable results with varying degrees of constraints. Meanwhile, the larger the alt limit, the smaller the standard deviation. The distributions of airmass and the limiting magnitude show that observations around the zenith will be prioritized.

To better explore the effects of the observation time, site distribution, and field distribution on the scheduling results, and testify the robustness of the proposed method, a series of

---

[7] https://github.com/lsst/rubin_sim

Table 3
Median and Standard Deviations of the Airmass and $5\sigma$ Limiting Magnitude, and Four Metrics (UNO, UCOT, UC, and TAR) under Different Alt Constraints after the Duration of 10 Long-term Scheduling Processes

|  | Alt > 0 deg | Alt > 30 deg | Alt > 60 deg |
| --- | --- | --- | --- |
| Airmass | 1.37, 0.32 | 1.30, 0.25 | 1.07, 0.04 |
| Limiting Magnitude | 23.54, 0.30 | 23.59, 0.27 | 23.78, 0.17 |
| UNO | 2.65 | 2.01 | 2.48 |
| UCOT | 13.25 | 10.03 | 12.41 |
| UC | 37.56 | 30.53 | 27.89 |
| TAR | 93.50% | 92.89% | 84.44% |

**Note.** There is not much difference in the scheduling metrics. The observations closer to the zenith can obtain better observation quality.

ablation experiments were conducted using data sets of 10 observation sites and 200 fields. More details and the results of the experiments are shown in Appendices A, B, and C.

## 6. Conclusion and Future Work

The coming decade will see new imaging and spectroscopy surveys of unprecedented scales on the ground and in space. Collaborative observations with telescope arrays lead to new modes of survey observation, facilitating more frequent and more extensive observations. Unlike mainstream telescope schedulers, the schedulers of telescope arrays do not rely on handcrafted observation proposals. Instead, they require automatic operation and efficient control. Several constraints such as resource requirements, astronomical observation conditions, and real-time weather need to be considered, and they may compete with each other and change dynamically over time.

This paper proposes a multilevel scheduling model and implements a flexible framework that can successfully solve a generic version of the telescope-array scheduling problem considering the projected volumes of constraints and objectives. The modular design and scalability of our method provide a desirable environment and feasible solution for a wide-range of programming expertize in the astronomy community. This property allows for expert intervention if needed by simple parameter adjusting of the configuration file to adapt to specific projects. Also, by integrating the recently proposed novel solvers, it can easily support further performance optimization.

This scheduling algorithm features a multilayer model that performs a schedule process in both the global-level and site-level phases. Also, a time discretization and both static and dynamic variables are considered. In addition, the proposed framework offers a systematic approach to the optimization of the schedulers behavior under uncertainties and interruptions, and can support rescheduling if needed due to changes that may affect the scheduling process. The functions and performances of the scheduler are evaluated through simulated scenarios.

Experimental results reveal that our method achieves acceptable performance according to the obtained percentages of the time allocation efficiency and the uniformity of sky coverage. Additionally, the established parameter configuration is able to solve all simulated instances in a feasible amount of time without using any sophisticated hardware environment. Compared with existing single-telescope time-domain survey projects like ZTF and LSST, the quality of the observed images in the experimental simulated telescope array is more difficult to be guaranteed due to more complex observation constraints and objectives. Experimental results show that our approach works well in different situations (e.g., different constraints, observation time, various distributions of observation sites/fields).

In addition, our experiments find that for telescope-array observation scheduling, there is a trade-off problem between ensuring the uniformity of the sky survey and enhancing the quality of single observation images, and it is necessary to consider the balance between these multiple optimization objectives in a more refined and comprehensive way. Alternatively, the scheduler can be oriented to the characteristics of the different observation modes and optimization targets and be classified and customized to achieve faster response and better efficiency. Multiple mode switching and collaborative response are also issues worth exploring in the future.

We suggest that the proposed formalism and framework would provide useful clarity to similar problems and applications in operation research and computer science, such as the satellite scheduling problem. In future research, we will further refine and expand our framework to include the task scheduling problem of each telescope in a shorter period in second- and third-level scheduling, considering the exposure time of each site's telescopes and the filter requirements of the specific observation project. Moreover, we will improve our algorithm and continue researching more efficient scheduling algorithms for telescope arrays to which there is still a huge field to contribute. To fully reap the scientific value of the large-scale surveys and investments, we hope astronomers give sustained attention to the scheduling problems of telescope arrays.

This work was financially supported by the National Natural Science Foundation of China (NSFC; 12133010, 12273025, 11873010, 11733007), and the Nebula Talents Program of the National Astronomical Observatories, CAS.

## Appendix A
## Effect of the Observation Time

To investigate the effect of the input observation time on the scheduling results, three main scenarios are used for testing. One of the important time-varying factors influencing the observations is the lunar phase. So the three scenarios observe the same fields in a month (the moon phase details are shown in Figure 10), use the same distribution of observation sites, but set different lunar illumination limits, (1) taken with less than 5% of the lunar illumination, (2) moon up and illuminated from 5% to 50%, and (3) moon up illuminated more than 50%.

Figure 11 and Table 4 show the airmass and $5\sigma$ limiting magnitude results and the results of the proposed four metrics (UNO, UCOT, UC, and TAR). It can be seen that the trends of the airmass and limiting magnitude results are quite similar for the different scenarios. In addition, because a telescope has different visible times for a target observation field with different lunar illumination limits, the scheduling results are different (Figure 12). The results of observation cadence uniformity and telescope time utilization reflect that, when more observation time slots are available, the scheduling algorithm will sacrifice part of the telescope time utilization to consider more observation uniformity among fields.

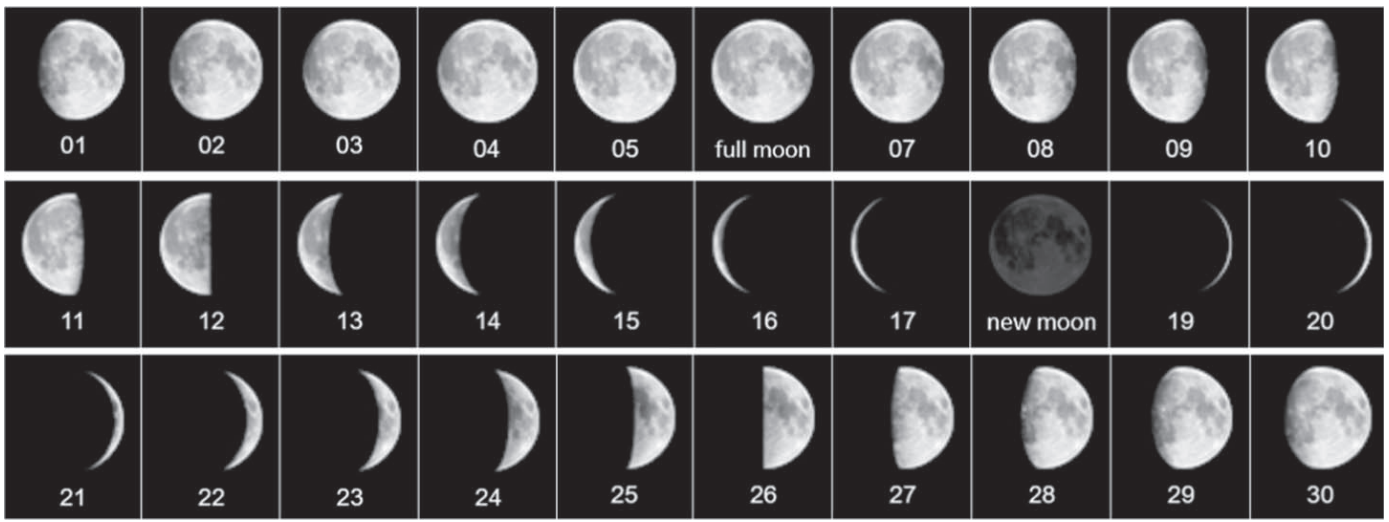

**Figure 10.** Schematic diagram of the lunar phase change of the one-month observation time in the experiments.

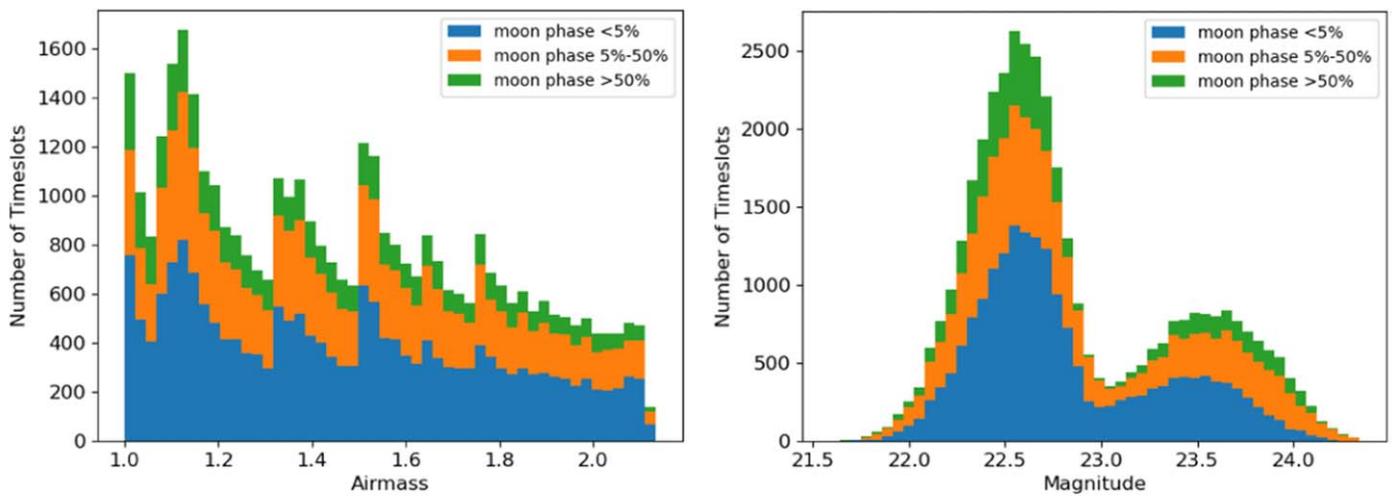

**Figure 11.** Histograms of the airmass and 5σ limiting magnitude results for scenario 1 (moon phase <5%), scenario 2 (moon phase 5%–50%), and scenario 3 (moon phase > 50%), after the duration of a one-month scheduling process.

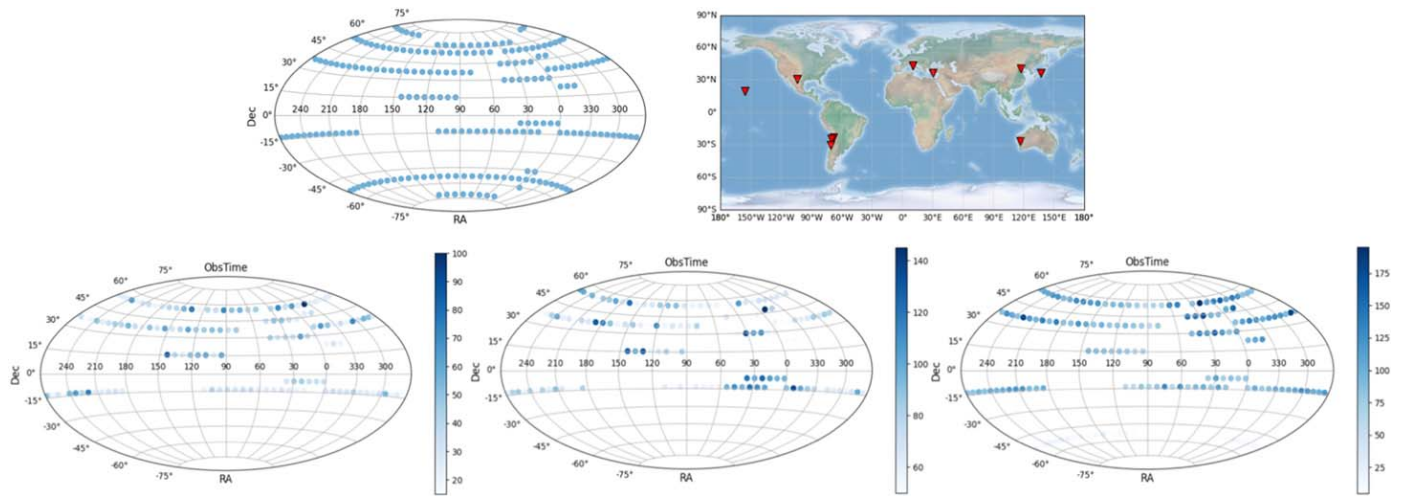

**Figure 12.** The distributions of the fields and the observation sites used are illustrated in the first row. In the second row, we show the coverage of the cumulative observation time (in minutes) for scenario 1 (left), scenario 2 (middle), and scenario 3 (right) after the duration of a one-month scheduling process.

Table 4
Median and Standard Deviations of the Airmass and 5σ Limiting Magnitude, and Four Metrics (UNO, UCOT, UC, and TAR) for Scenario 1, Scenario 2, and Scenario 3

|  | Scen1 | Scen2 | Scen3 |
| --- | --- | --- | --- |
| Airmass | 1.44, 0.32 | 1.47, 0.31 | 1.46, 0.31 |
| Limiting Magnitude | 22.84, 0.58 | 22.90, 0.60 | 22.81, 0.49 |
| UNO | 3.13 | 4.68 | 8.25 |
| UCOT | 15.67 | 23.39 | 41.28 |
| UC | 215.72 | 168.68 | 158.82 |
| TAR | 90.97% | 76.30% | 68.30% |

## Appendix B
## Effect of the Site Distribution

To examine how the input observation sites affect the final scheduling results, we also evaluate schedules with different distributions of sites. Scenario 4 is generated using more observation sites distributed in the northern hemisphere while scenario 5 uses more telescopes in the southern hemisphere. See Figure 13 for the geographical distributions of the observation sites. The observation time and the fields used for the scheduling calculation of scenario 4 and scenario 5 are the same, and the obtained experimental results are shown in Figures 13, 14, and Table 5.

In the simulated scenarios, part of the input fields is unobservable during the calculated observation time due to their relative position to the telescopes (Figure 14), which also causes the fluctuations of UC and TAR shown in Table 5. The observation quality represented by the airmass and limiting magnitude is relatively stable, suggesting that our approach could well adapt to different site distributions.

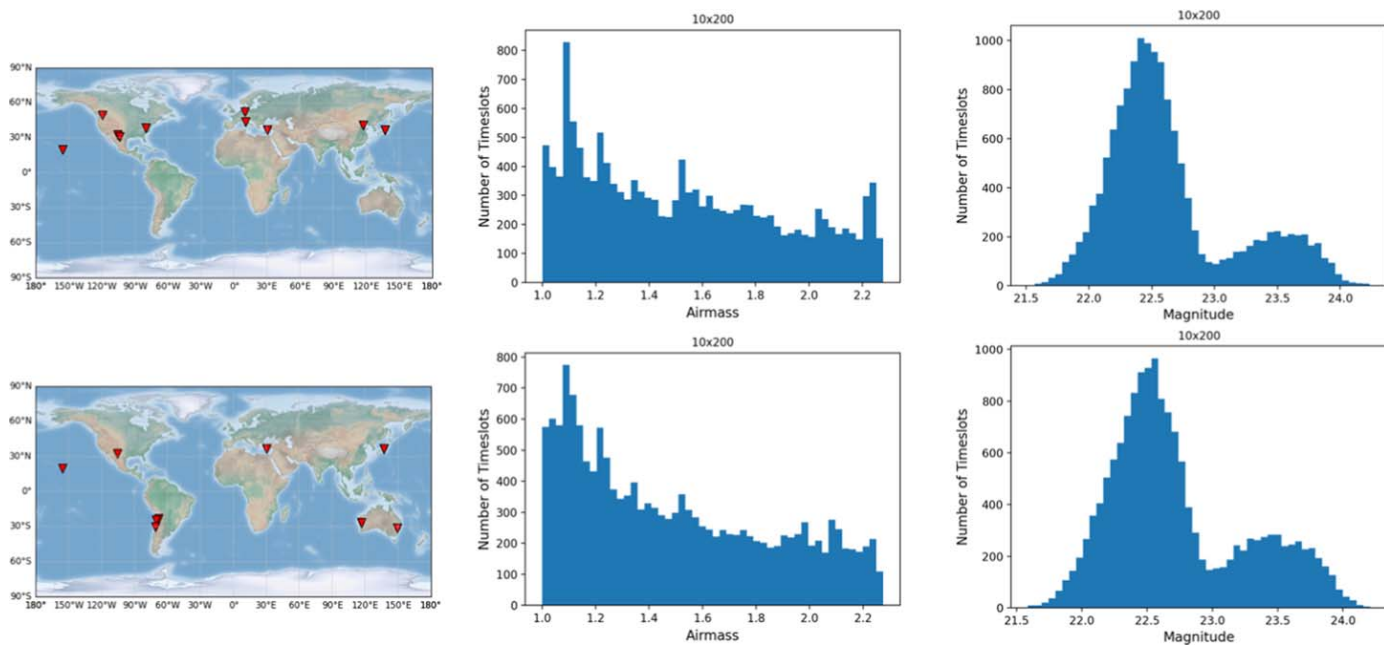

Figure 13. Distribution of the input observation sites and airmass and 5σ limiting magnitude results after the same duration of 10 long-term scheduling processes for scenario 4 (first row) and scenario 5 (second row).

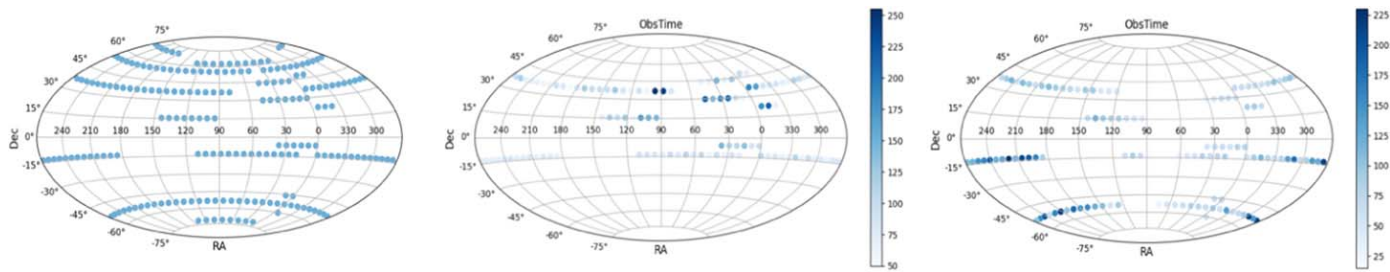

Figure 14. Left: distribution of input fields. Middle and right: coverage of the cumulative observation time (in minutes) for scenario 4 and scenario 5, respectively, after the same duration of 10 long-term scheduling processes.

Table 5
Median and Standard Deviations of the Airmass and 5σ Limiting Magnitude, and Four Metrics (UNO, UCOT, UC, and TAR) for Scenario 4 and Scenario 5

|                    | Scen4        | Scen5        |
|--------------------|--------------|--------------|
| Airmass            | 1.53, 0.37   | 1.49, 0.37   |
| Limiting Magnitude | 22.64, 0.51  | 22.73, 0.52  |
| UNO                | 8.29         | 9.26         |
| UCOT               | 41.46        | 46.31        |
| UC                 | 72.40        | 32.48        |
| TAR                | 69.18%       | 80.71%       |

## Appendix C
## Effect of the Field Distribution

We also evaluated the impact of the field distribution on the scheduling results for the same input observation time. The same number of fields are generated but distributed at low ([−30°, 30°]), middle ([−60°, −30°] and [30°, 60°]), and high latitudes ([−90°, −60°]), represented by scenario 7, scenario 8, and scenario 9, respectively. In particular, considering that telescopes used to observe low- and middle-latitude fields have difficulties observing high-latitude regions, we select five sites that can observe high-latitude fields and assume that double the number of sites are used at each site location to keep the total number of sites consistent. See details in Figure 15.

We find that, although the available observation time is different (Figure 16), the overall observation quality and uniformity are not too much affected by the observation field variation (Figure 15 and Table 6). For high-latitude observation fields, note that due to the limitation in the selected observation equipment positions, the airmass of the scheduled observation is relatively high, the value of the limiting magnitude is low, and thus only one wave peak appears in Figure 15. At the same time, as the locations of the fields set in scenario 8 are more concentrated, the value of UC is also smaller, which is consistent with our expectation. Therefore, this demonstrates that when the telescopes can provide normal observation conditions, our scheduling method is able to obtain stable scheduling results regardless of the distribution of the input fields.

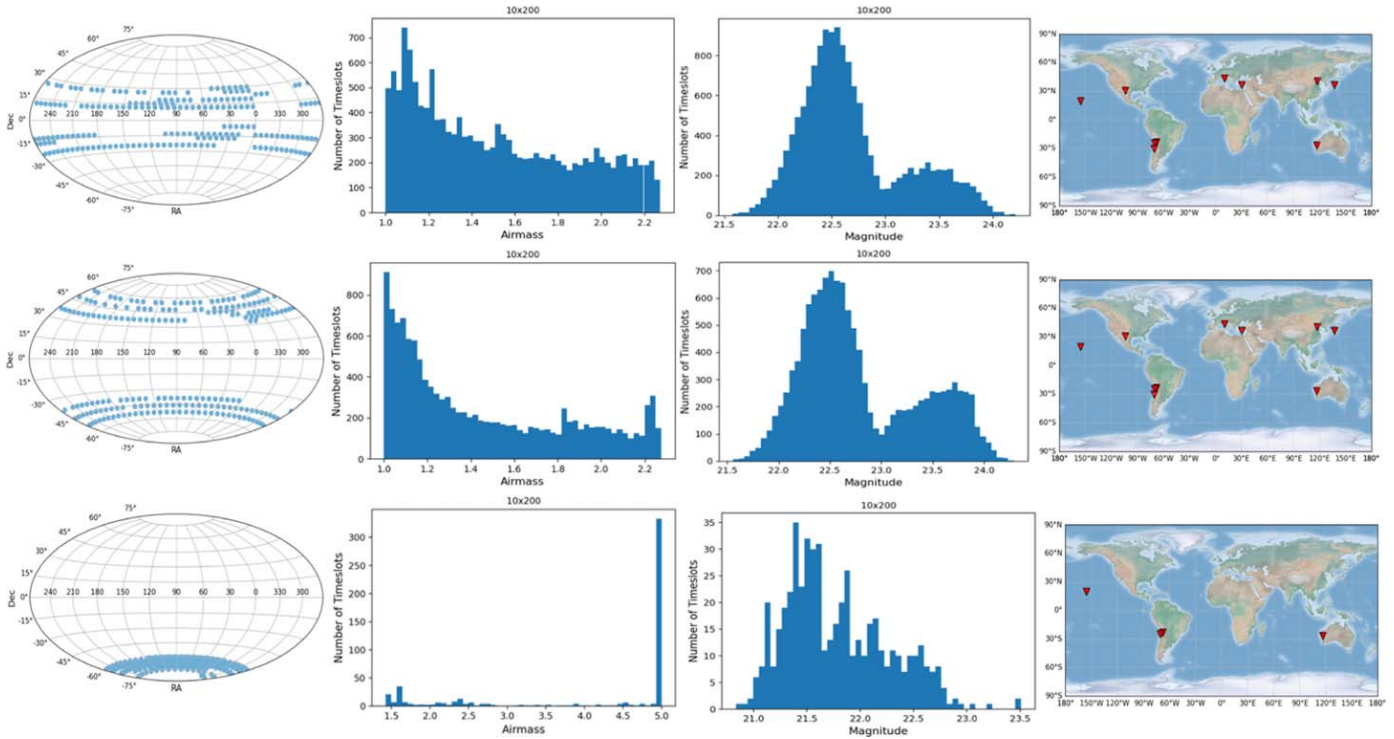

Figure 15. Histograms of the airmass and 5σ limiting magnitude results for scenario 6 (first row), scenario 7 (second row), and scenario 8 (third row), after the same duration of 10 long-term scheduling processes. The distributions of the fields and the observation sites are on the left and right sides, respectively.

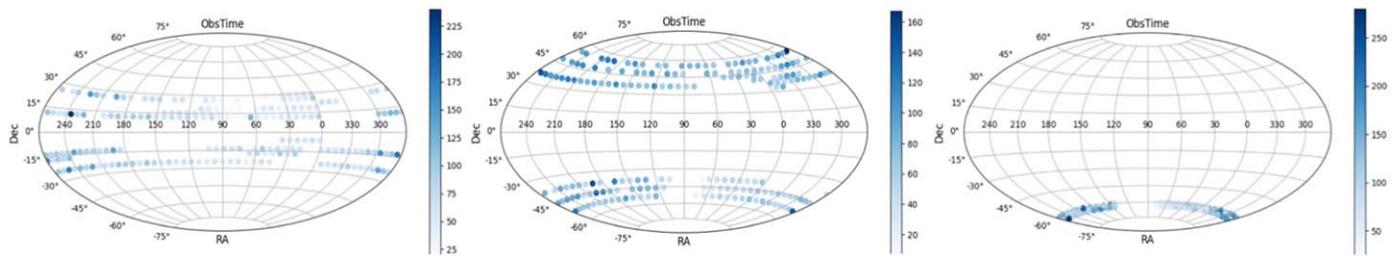

**Figure 16.** Coverage of the cumulative observation time (in minutes) for scenario 6 (left), scenario 7 (middle), and scenario 8 (right) after the same duration of 10 long-term scheduling processes.

**Table 6**
Median and Standard Deviations of the Airmass and $5\sigma$ Limiting Magnitude Distribution, and Four Metrics (UNO, UCOT, UC, and TAR) for Scenario 6, Scenario 7, and Scenario 8

|  | Scen6 | Scen7 | Scen8 |
| --- | --- | --- | --- |
| Airmass | 1.53, 0.37 | 1.45, 0.39 | 4.10, 1.36 |
| Limiting Magnitude | 22.76, 0.57 | 22.79, 0.58 | 21.78, 0.48 |
| UNO | 5.33 | 6.30 | 10.66 |
| UCOT | 26.67 | 31.51 | 53.29 |
| UC | 151.73 | 175.86 | 42.42 |
| TAR | 74.71% | 65.71% | 63.89% |

### ORCID iDs

Yajie Zhang  https://orcid.org/0000-0003-2976-8198
Yi Hu  https://orcid.org/0000-0003-3317-4771

### References

Bellm, E. C., Kulkarni, S. R., Graham, M. J., et al. 2018, PASP, 131, 018002
Alexander, K. D., Berger, E., Bower, G., et al. 2017, arXiv:1703.04692
Bellm, E. C., Kulkarni, S. R., Barlow, T., et al. 2019, PASP, 131, 68003
Bellman, R. 1966, Sci, 153, 34
Castro-Tirado, A. J., Moreno, F. S., del Pulgar, C. P., et al. 2014, RMxAA, 45, 104
Chen, X., & Tian, Y. 2019, in Advances in Neural Information Processing Systems, Vol. 32 (Red Hook, NY: Curran Associates, Inc.), https://proceedings.neurips.cc/paper/2019/file/131f383b434fdf48079bff1e44e2d9a5-Paper.pdf
Colome, J., Colomer, P., Guàrdia, J., et al. 2012, Proc. SPIE, 8448, 84481L
Dyer, M. J. 2020, arXiv:2003.06317
Floudas, C. A., & Lin, X. 2005, Ann. Oper. Res., 139, 131
Gao, K., Cao, Z., Zhang, L., et al. 2019, IEEE/CAA J. Autom. Sin., 6, 904
Ghasemi, A., Ashoori, A., & Heavey, C. 2021, Appl. Soft Comput., 106, 107309
Graham, M. J., Kulkarni, S. R., Bellm, E. C., et al. 2019, PASP, 131, 078001
Gurobi Optimization, Inc. 2014, Gurobi Optimizer Reference Manual, http://www.gurobi.com
Hunter, J. D. 2007, CSE, 9, 90
Ivezić, Ž., Kahn, S. M., Tyson, J. A., et al. 2019, ApJ, 873, 111
Lampoudi, S., Saunders, E., & Eastman, J. 2015, in ICORES 2015-4th Int. Conf. on Operations Research and Enterprise Systems, Proc. (Lisbon: SciTePress), 331
LIU, J., Soria, R., Wu, X.-F., Wu, H., & Shang, Z. 2021, Anais da Academia Brasileira de Ciências, 93, 20200628
Liu, Q., Wei, P., Shang, Z.-H., Ma, B., & Hu, Y. 2018, RAA, 18, 005
López-Casado, C., Pérez-del Pulgar, C., Mu noz, V. F., & Castro-Tirado, A. J. 2019, Future Gener. Comput. Syst., 95, 116
Morris, B. M., Tollerud, E., Sipőcz, B., et al. 2018, AJ, 155, 128
Naghib, E., Yoachim, P., Vanderbei, R. J., Connolly, A. J., & Jones, R. L. 2019, AJ, 157, 151
Price-Whelan, A. M., Sipőcz, B., Günther, H., et al. 2018, AJ, 156, 123
Sol 2016, A&C, 15, 90
Van Der Walt, S., Colbert, S. C., & Varoquaux, G. 2011, CSE, 13, 22
Wootten, A. 2003, Large Ground-based Telescopes, 4837, 110
Yokoo, M., Durfee, E. H., Ishida, T., & Kuwabara, K. 1998, IEEE Trans. Knowl. Data Eng., 10, 673